\documentclass
[runningheads]{llncs}
\usepackage[T1]{fontenc}
\usepackage{float}
\usepackage{chngcntr}
\usepackage{graphicx}
\usepackage{hyperref}
\usepackage{booktabs}
\usepackage{listings}
\usepackage{subcaption}
\usepackage{amsmath} 
\usepackage{tikz}
\usepackage{todonotes}
\usetikzlibrary{patterns, positioning}

\newcommand{\orcid}[1]{\href{https://orcid.org/#1}{\textcolor[HTML]{A6CE39}{\aiOrcid}}}

\begin{document}


\title{Towards a Scalable and Efficient PGAS-based Distributed OpenMP}

\titlerunning{PGAS-based Distributed OpenMP}
%
\author{Baodi Shan\inst{1} \and
Mauricio Araya-Polo\inst{2} \and
Barbara Chapman\inst{1}}
\authorrunning{Shan et al.}
%
\institute{Stony Brook University, Stony Brook NY 11794, USA \\
\email{\{baodi.shan,barbara.chapman\}@stonybrook.edu}
\and
TotalEnergies EP Research \& Technology US, LLC, Houston TX 77002, USA\\
}
\maketitle              
\begin{abstract}

MPI+X has been the \textit{de facto} standard for distributed memory parallel programming. It is widely used primarily as an explicit two-sided communication model, which often leads to complex and error-prone code. Alternatively, PGAS model utilizes efficient one-sided communication and more intuitive communication primitives.
In this paper, we present a novel approach that integrates PGAS concepts into the OpenMP programming model, leveraging the LLVM compiler infrastructure and the GASNet-EX communication library.
Our model addresses the complexity associated with traditional MPI+OpenMP programming models while ensuring excellent performance and scalability.
We evaluate our approach using a set of micro-benchmarks and application kernels on two distinct platforms: Ookami from Stony Brook University and NERSC Perlmutter. The results demonstrate that DiOMP achieves superior bandwidth and lower latency compared to MPI+OpenMP, up to $25\%$ higher bandwidth and down to $45\%$ on latency. 
DiOMP offers a promising alternative to the traditional MPI+OpenMP hybrid programming model, towards providing a more productive and efficient way to develop high-performance parallel applications for distributed memory systems.

\end{abstract}

\section{Introduction}
%
%

HPC systems continue to grow in size and complexity, pushing legacy programming models to their limits. Developers of numerical simulation applications must adapt to this reality. Fortunately, alternative programming models and productivity frameworks are available and continually evolving to provide necessary support. Currently and for most of the last decade, MPI+X is the mainstream paradigm for distributed cluster programming models, where X can be OpenMP, OpenACC, CUDA, RAJA or Kokkos, etc~\cite{MPIOpenMP,MPIKokkos,MPICUDA}. However, there is an increasing need for alternatives to MPI+X that are more flexible and less complex. One such alternative is the PGAS (Partitioned Global Address Space) programming model, which is gaining momentum. Notable PGAS models such as UPC++, OpenSHMEM, and Legion and languages such as Chapel are reaching larger developer audiences.

OpenMP is rapidly evolving from a traditional CPU-based and shared-memory programming model to one that includes task-based programming and accelerator-based offloading capabilities. Therefore, we aim to leverage the power of PGAS to extend OpenMP to operate in distributed environments. To that end, we propose the PGAS-based Distributed OpenMP (\textbf{DiOMP}). DiOMP's main contributions are:

\textbf{Enhanced Scalability and Improved Performance:}
DiOMP boosts performance and scalability for distributed applications by allowing efficient data sharing across nodes without the overhead of traditional message-passing.

\textbf{Simplified Communication in the PGAS Model:}
DiOMP exploits the PGAS model direct operations on global memory addresses, which reduces the complexities of message matching and buffer management commonly found in MPI. In the PGAS framework, communication operations like reading and writing remote data are conducted directly via global addresses, without the need for additional management of communication domains. 



\textbf{Simplified Memory Management:}
By extending native OpenMP statements, such as \texttt{omp\_alloc()}, this model simplifies the allocation and management of memory. Compared to MPI RMA, this approach avoids the complexities and overhead associated with creating and destroying MPI windows. 

\textbf{Excellent Extensibility through Activate Message:}
Active Messages is a communication mechanism that reduces latency and overhead by directly executing a handler function upon message arrival, ensuring efficient and immediate processing. This guarantees the extensibility of DiOMP, in the current version of DiOMP, \texttt{ompx\_lock()} is implemented using Active Messages. In future versions, Active Messages will play a crucial role in handling task dependencies within DiOMP by allowing for dynamic and responsive communication patterns.

\section{Background}

\subsection{OpenMP}

OpenMP~\cite{openmp5} is one the main standard for shared-memory parallelism in HPC. It provides a straightforward and flexible interface for developers to create parallel applications by exploiting the capabilities of multi-core processors and shared memory systems.
Current versions of OpenMP support the task-based programming model\cite{shileitask}, for instance, OpenMP 4.0 introduced task dependencies, allowing programmers to specify dependencies between tasks and enabling the runtime system to automatically manage the execution order based on these dependencies. 
With the introduction of version 4.0, OpenMP also expanded its capabilities to include device offloading\cite{arxivbaodi}, enabling code execution on accelerators without requiring users to develop device-specific kernels using vendor-specific APIs\cite{9741290,9820621}.

\subsection{The PGAS Model}

PGAS stand for Partitioned Global Address Space programming model.
In contrast to the message-passing model (MPI), the PGAS programming model~\cite{pgasintro} utilizes a globally accessible memory space that is divided among the basic units distributed across one or more nodes. 

PGAS models offer a uniform view of distributed memory objects and enable high-performance access to remote memory through direct operations such as reads (\textbf{get}s) and writes (\textbf{put}s). Point-to-point communication in the PGAS model is one-sided, requiring active participation only from the initiating unit. 
This decouples communication and synchronization, allowing the target unit's computation to continue uninterrupted during data exchanges. 

Many distributed and parallel computing programming languages and libraries feature the PGAS model, including OpenSHMEM, Legion, UPC++, DASH, Chapel, and OpenUH Co-Array Fortran. In the programming languages and libraries that have adopted PGAS, some use MPI as their communication framework, such as DASH, while others utilize UCX, such as OpenSHMEM. But the de-facto communication standard targeted by portable PGAS system is GASNet API. Current and historical GASNet clients include: UPC++~\cite{upcxx}, Cray Chapel~\cite{chapel}, Legion~\cite{legion}, OpenUH Co-Array Fortran~\cite{openuh}, OpenSHMEM Reference implementation~\cite{openshmem}, Omni XcalableMP~\cite{omni}, and several miscellaneous projects.

\subsection{Related Work}
\label{sec:rel}

The idea of executing OpenMP programs within distributed architectures has been extensively explored in scholarly research. The concept of Remote OpenMP offloading, as introduced by Patel and Doerfert~\cite{remote}, together with subsequent enhancements~\cite{optremote,mpiremote} and practical implementations, has demonstrated considerable promise for facilitating OpenMP target offloading to remote devices. Nonetheless, as noted in reference~\cite{optremote}, the scalability of such remote offloading is sub-par when compared with conventional hybrid MPI+OpenMP methodologies. In a similar line of analysis, the OpenMP Cluster developed by Yviquel et. al~\cite{OMPC}, which also focuses on OpenMP target offloading, conceptualizes remote nodes as a computational resource for OpenMP targets. Another path to distributed directive-based programming approach is by combining XMP and YML~\cite{petiton2014multi}. 


\section{Design of PGAS-based Distributed OpenMP}

PGAS-based Distributed OpenMP is developed based on LLVM/OpenMP and utilizes GASNet-EX as the underlying communication middleware. In this section, we will sequentially introduce the memory management model of our PGAS-based approach, point-to-point communication, collective communication, and the synchronization mechanisms, as well as the role and future potential of GASNet-EX Active Message in these mechanisms.

\subsection{Memory Management}

\begin{figure}[h]
    \centering
    \includegraphics[width=\textwidth]{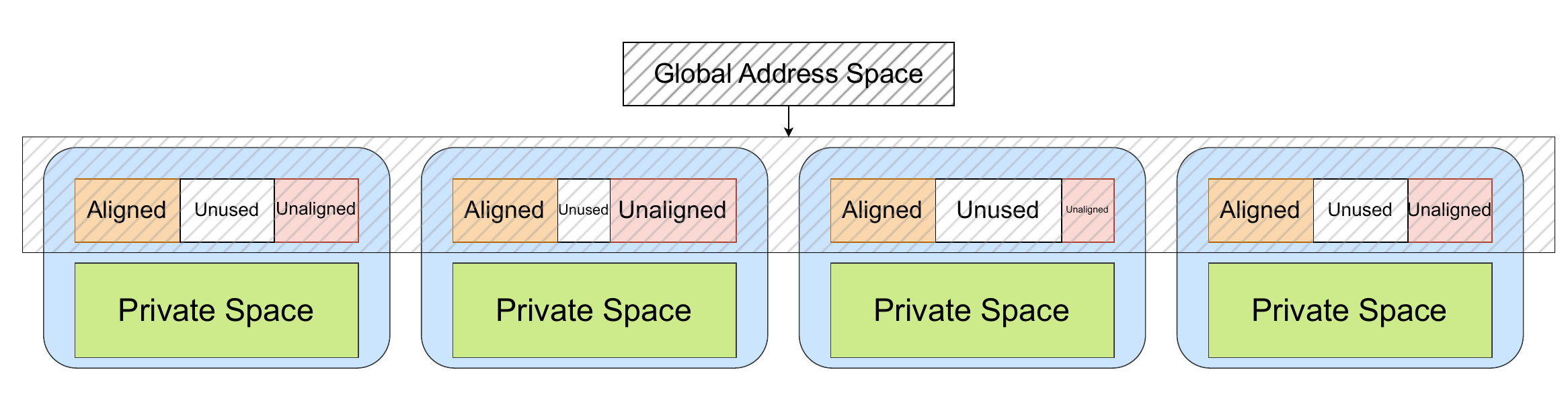}
    \caption{Memory management model of the DiOMP. Each node has its own private space (green) and a shared global space (striped), with the global space further divided into aligned space (orange) and unaligned space (red). The white parts represent unused (unallocated) memory space.}
    \label{fig:pgas}
\end{figure}

In the PGAS layer, we use process (\textit{rank}) as the main unit for memory management and communication. The memory region of each rank is divided into private memory and global memory, adhering to the PGAS paradigm guidelines. The memory management model is illustrated in~\autoref{fig:pgas}.
Due to the segment constraints imposed by the communication middleware GASNet-EX, the memory space related to communication must reside within the segment previously allocated via \texttt{gex\_Segment\_Attach()}. To address this requirement, we introduce aligned global memory and unaligned global memory. The allocation of aligned global memory needs the involvement of all ranks, with each rank acquiring an equal size of global memory, which is then placed at the front of their respective segments.

The segments attached by GASNet-EX do not support address alignment, meaning that GASNet-EX cannot guarantee identical address ranges across different ranks' segments. Therefore, PGAS-based distributed OpenMP uses virtual address alignment.
Virtual address alignment operates as follows: although the actual memory addresses assigned to each rank during the allocation of aligned global memory may differ, the runtime system maintains a specific mapping that provides a virtually aligned address space. Thus, when a rank intends to transfer data to other ranks, it can simply utilize its own memory address to obtain the corresponding memory addresses of the other ranks.


As for non-aligned global memory, which is global memory that can be created by individual or some ranks. This type of memory is allocated at the end of the segment in a limited manner. This memory does not receive virtual address mapping, as the process is invisible to ranks that do not participate in this portion of the memory allocation. Non-aligned global memory is particularly suitable for storing and retrieving specific or temporary data.

Whether using aligned or non-aligned global memory, developers utilizing DiOMP can easily allocate global memory by invoking the \texttt{omp\_alloc()} function, which is part of the OpenMP standard. DiOMP is equipped with specially designed allocators for allocating data in the global space. In addition to supporting standard \textbf{C} function from OpenMP, we have also provided a \textbf{C++} allocation function with template support, enabling developers to allocate memory for specific data type or data structure.

\subsection{Point-to-Point and Collective Communication}

DiOMP incorporates two fundamental communication paradigms: point-to-point and collective communication. These paradigms enhance data exchange and synchronization across different ranks, facilitating efficient parallel computing.
\lstset{
    language=C,             
    basicstyle=\fontsize{9}{12}\ttfamily ,
    numbers=left,           
    numberstyle=\tiny,      
    stepnumber=1,           
    numbersep=5pt,          
    backgroundcolor=\color{white}, 
    showspaces=false,       
    showstringspaces=false, 
    showtabs=false,         
    frame=single,           
    captionpos=b,           
    breaklines=true,        
    breakatwhitespace=false, 
    title=\lstname          
}

\counterwithout{lstlisting}{chapter}  
\renewcommand{\thelstlisting}{\arabic{lstlisting}}
\begin{lstlisting}[caption={Point-to-point APIs for PGAS-based Distributed OpenMP}, label=lst:rma]
void ompx_get(void *dst, int rank, void *src, size_t nbytes);
void ompx_put(int rank, void *dst, void *src, size_t nbytes);
\end{lstlisting}
\setcounter{footnote}{0}
\renewcommand*{\thefootnote}{\fnsymbol{footnote}}

Point-to-point communication leverages one-sided communication primitives, including \texttt{put} and \texttt{get}\footnote{The model and framework proposed in this paper are currently limited to the proof of concept stage, and the function names are provisional.}. This method enables ranks to directly access each other's memory without needing explicit coordination, thus reducing synchronization overhead and allowing computation and communication to overlap. These operations could utilize a virtual address alignment mechanism to seamlessly map between local and remote memory spaces.~\autoref{lst:rma} shows the APIs for point-to-point communication in DiOMP. 
Collective communication, on the other hand, requires all ranks to participate in data exchange or synchronization. DiOMP supports various collective operations like barrier, broadcast, and reduction, which are optimized based on the network topology and hardware capabilities. These operations help in the efficient distribution and aggregation of data, supporting common parallel programming patterns. 
Together, these communication strategies provide a robust framework in PGAS-based Distributed OpenMP. 

\subsection{Synchronization Mechanisms and Active Messages}

DiOMP based on GASNet-EX offers a variety of synchronization mechanisms, including \texttt{ompx\_barrier()}, \texttt{ompx\_waitRMA()}, and \texttt{ompx\_lock()}. Among these, the implementations of barrier and waitRMA are based on the native interfaces of GASNet-EX, while \texttt{ompx\_lock()} utilizes the Active Message mechanism of GASNet-EX. We will use \texttt{ompx\_lock()} as a case study to demonstrate the significant role that Active Message plays in our model.

\begin{figure}[h]
    \centering
    \includegraphics[width=.75\textwidth]{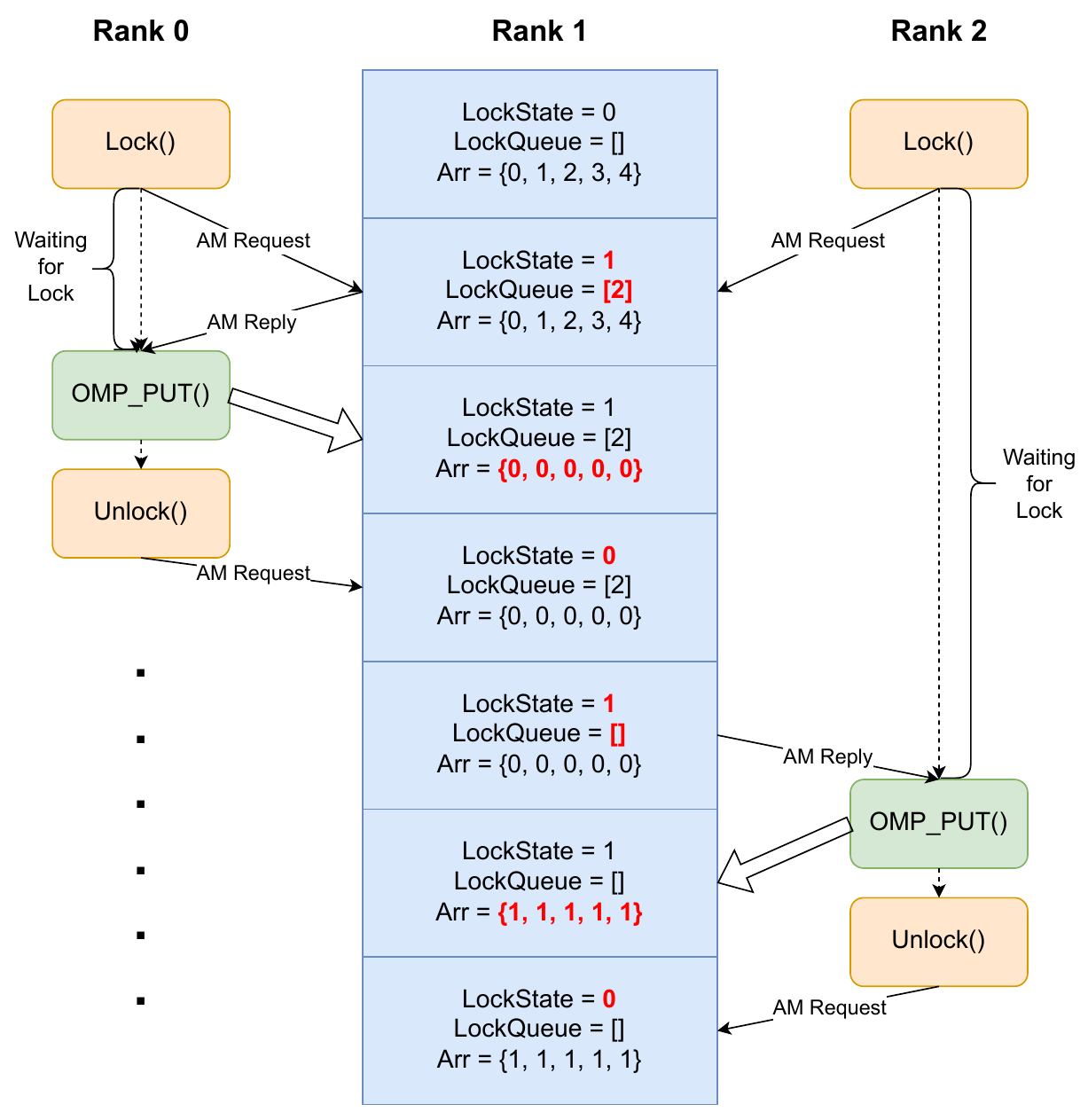}
    \caption{The workflow of the \texttt{ompx\_lock()} and \texttt{ompx\_unlock()} based on Active Messages in the presence of contention.}
    \label{fig:lock}
\end{figure}

The primary function of \texttt{ompx\_lock()} is to ensure that a specific rank has exclusive access to the shared memory space of a target rank by establishing a lock. This process is facilitated by several dedicated GASNet-EX active message handlers.
When one rank (source rank) wants to lock another rank (target rank), it starts by sending an active message. The source rank then waits for a reply to see if it got the lock. Meanwhile, the target rank checks this request and manages a list of all ranks waiting for a lock, along with a lock status indicator.
If no other rank is waiting for a lock and the target rank is not locked, the target rank will lock itself and inform the source rank that it has successfully obtained the lock through a reply active message. If the target rank is already locked or there are other ranks waiting, the source rank is added to the waiting list. The source rank must then wait its turn until it is at the front of the list and the target rank is unlocked.


Each active message handler in GASNet-EX possesses a unique token, which means the rank queue stores these tokens, each embodying information about its corresponding source rank. This mechanism ensures that every request is uniquely identified and correctly processed.
In cases where the lock cannot be immediately granted, the target rank does not idle. Instead, it monitors the rank queue and only responds once the locking rank issues an unlock active message. This efficient management prevents unnecessary delays and optimizes resource use.
\autoref{fig:lock} illustrates the process where rank0 and rank2 simultaneously initiate lock requests and put data on rank1. 

Building upon this, we have also introduced the \texttt{ompx\_lockt()} function, which is an extension of \texttt{ompx\_lock()} that provides thread-level locking.
This function implements both thread-level and process-level locking, making it extremely useful in mixed thread and process programming scenarios, such as when inter-rank communication occurs within an \texttt{omp parallel for} region.

In the future, we plan to further expand the role of active message within DiOMP, particularly in handling OpenMP task dependencies. Active message is expected to play a crucial role in this context.

\section{Evaluation}

\subsection{Experimental Setup}

The experiments were conducted on the Ookami system at Stony Brook University and the Perlmutter supercomputer at Lawrence Berkeley National Laboratory. Refer to~\autoref{tab:system} for the hardware and software specifications of the systems.
We performed micro-benchmarks on both systems and tested weak scaling matrix multiplication and strong scaling Minimod~\cite{meng2020minimod} benchmark on Ookami.

\begin{table}[htb]
\caption{Hardware and software configuration of the experimental platforms}
\centering
\renewcommand{\arraystretch}{1.5}
\begin{tabular}{l@{\hspace{5mm}}l@{\hspace{5mm}}l}
\hline
                      & \textbf{Ookami} & \textbf{Perlmutter} \\ \hline
\textbf{CPUs}         & Fujitsu A64FX   & AMD EPYC 7763 * 2   \\
\textbf{CPU cores}    & 48              & 64                  \\
\textbf{Memory}       & 32 GB HBM2      & 512 GB DDR4         \\
\textbf{Interconnect} & InfiniBand HDR  & HPE Slingshot-11    \\
\textbf{MPI}          & MVAPICH 2.3.7   & Cray MPICH 8.1.28   \\
\textbf{GASNet-EX}    & \multicolumn{2}
{c}{GASNet-2023.9.0}   \\ \hline
\end{tabular}
\label{tab:system}
\end{table}

\subsection{Micro-benchmarks}

We conducted micro benchmark tests on Ookami and Perlmutter platforms to evaluate the performance of DiOMP in terms of bandwidth and latency.

\begin{figure}[!h]
    \centering
    \includegraphics[width=\textwidth]{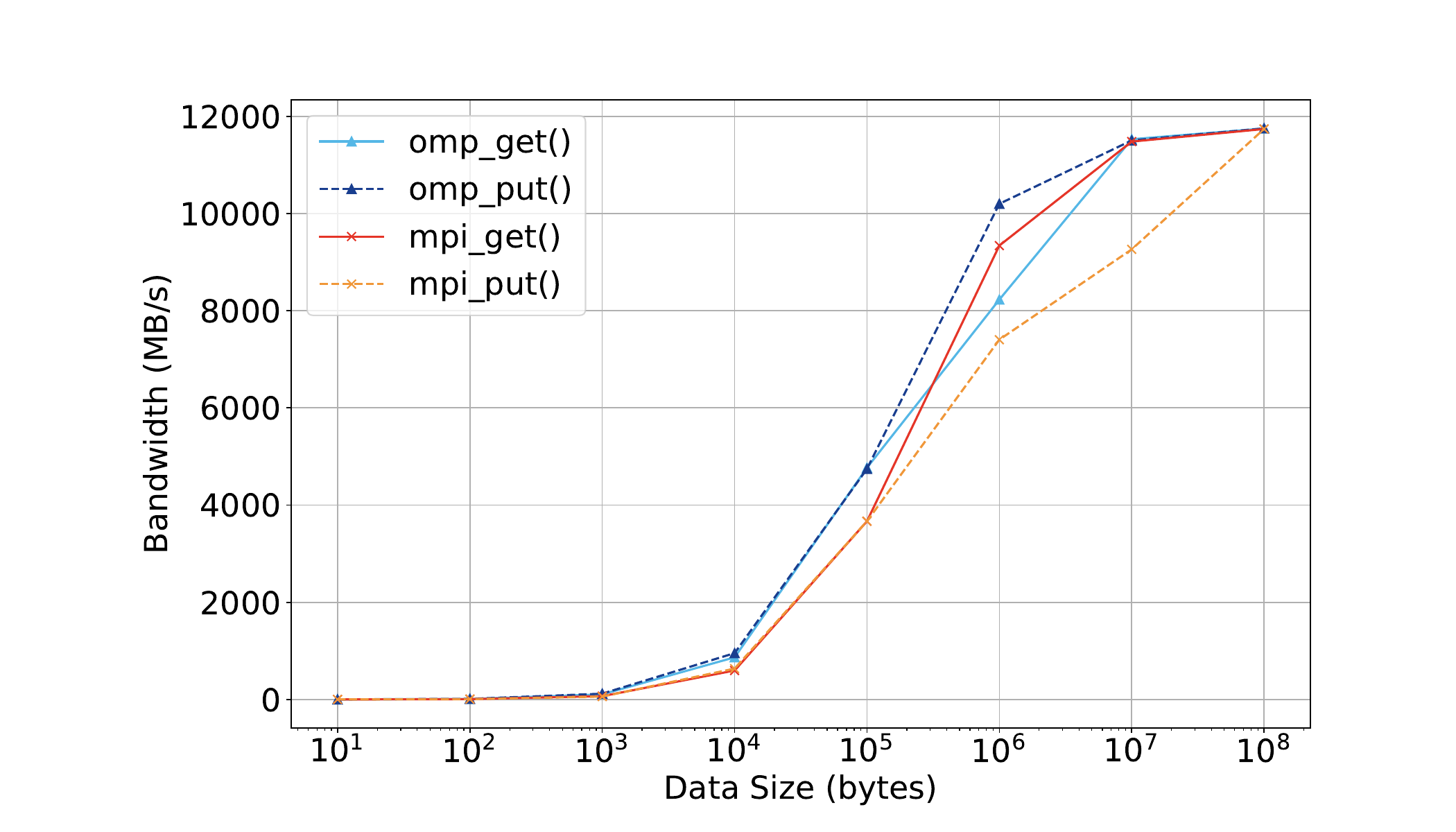}
   \caption{Micro-benchmark for bandwidth on Ookami}
    \label{fig:ookamibw}
\end{figure}

\begin{figure}[!h]
    \centering
    \includegraphics[width=\textwidth]{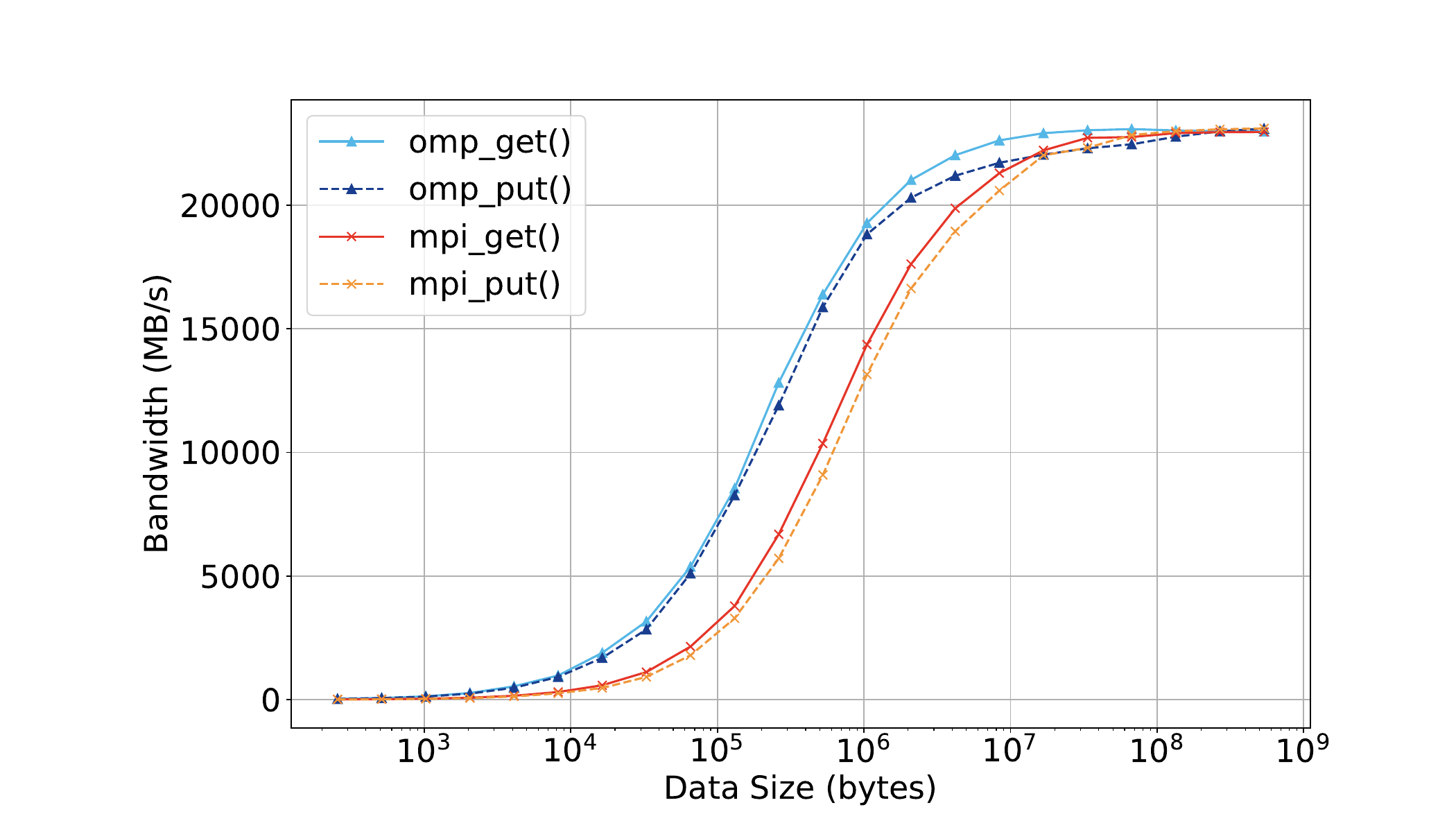}
    \caption{Micro-benchmark for bandwidth on Perlmutter. Notice that for messages of size $10^6$, PGAS+OpenMP outperforms MPI+OpenMP by $25\%$.}
    \label{fig:perlbw}
\end{figure}

The bandwidth tests using large message sizes showed that DiOMP achieved higher peak bandwidth and sustained higher throughput compared to MPI on both platforms (\autoref{fig:ookamibw} and \autoref{fig:perlbw}). As the message size increases, DiOMP-based implementation achieves peak bandwidth earlier than MPI. This can be attributed to the efficient utilization of the underlying interconnect through the GASNet-EX communication layer.

\begin{figure}[!htb]
    \centering
    \includegraphics[width=\textwidth]{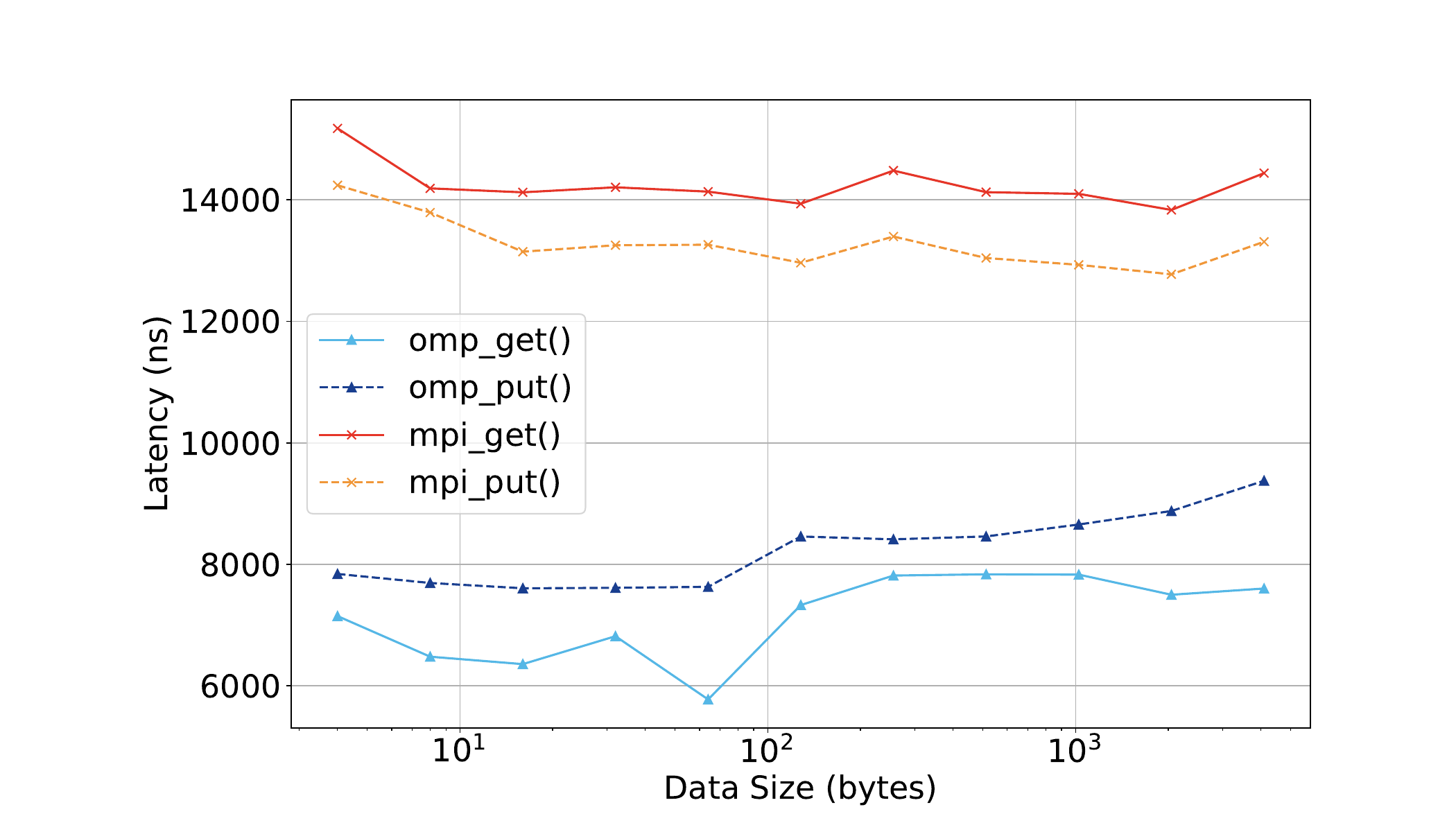}
    \caption{Micro-benchmark for latency on Ookami. Notice that PGAS+OpenMP latency across message sizes is in average $45\%$ lower then MPI+OpenMP.}
    \label{fig:ookamilat}
\end{figure}

\begin{figure}[!htb]
    \centering
    \includegraphics[width=\textwidth]{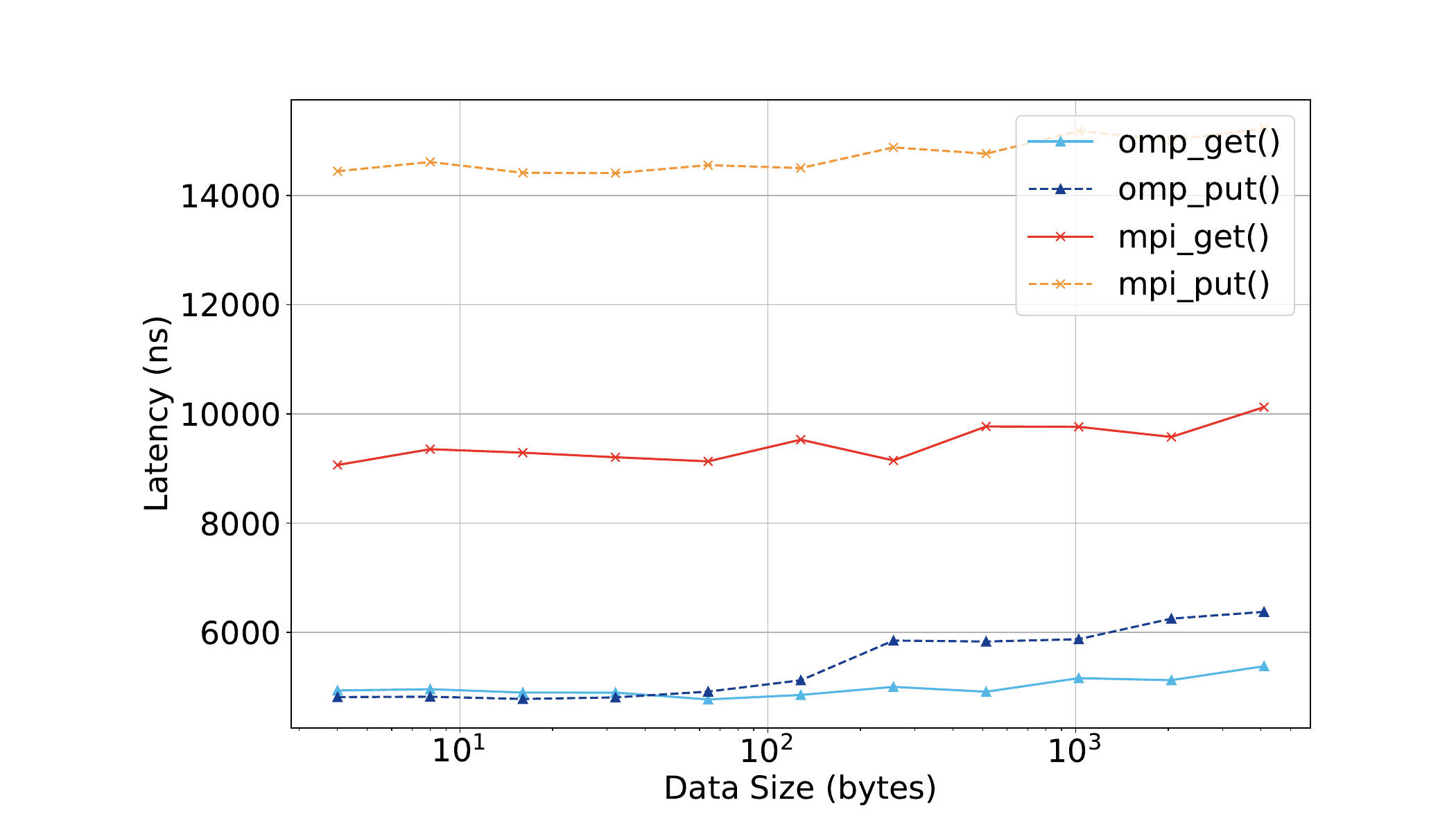}
    \caption{Micro-benchmark for latency on Perlmutter.}
    \label{fig:perlmutterlat}
\end{figure}

The latency tests using small message sizes demonstrated that DiOMP consistently demonstrates lower latency compared to MPI on both Ookami and Perlmutter (\autoref{fig:ookamilat} and \autoref{fig:perlmutterlat}). The reduction in latency is up to $45\%$. The lower latency of DiOMP is a result of its lightweight one-sided communication model, which eliminates the overhead associated with explicit message matching and synchronization in MPI. Notice that the performance of mpi\_put and mpi\_get on Perlmutter is consistent but apart, it has been previously reported~\cite{osti_1481769}.

These findings suggest that DiOMP is a promising alternative for high-performance inter-node communication in parallel applications.

\subsection{Weak Scaling - Matrix Multiplication}

\begin{figure}[!h]
    \centering
    \includegraphics[width=\textwidth]{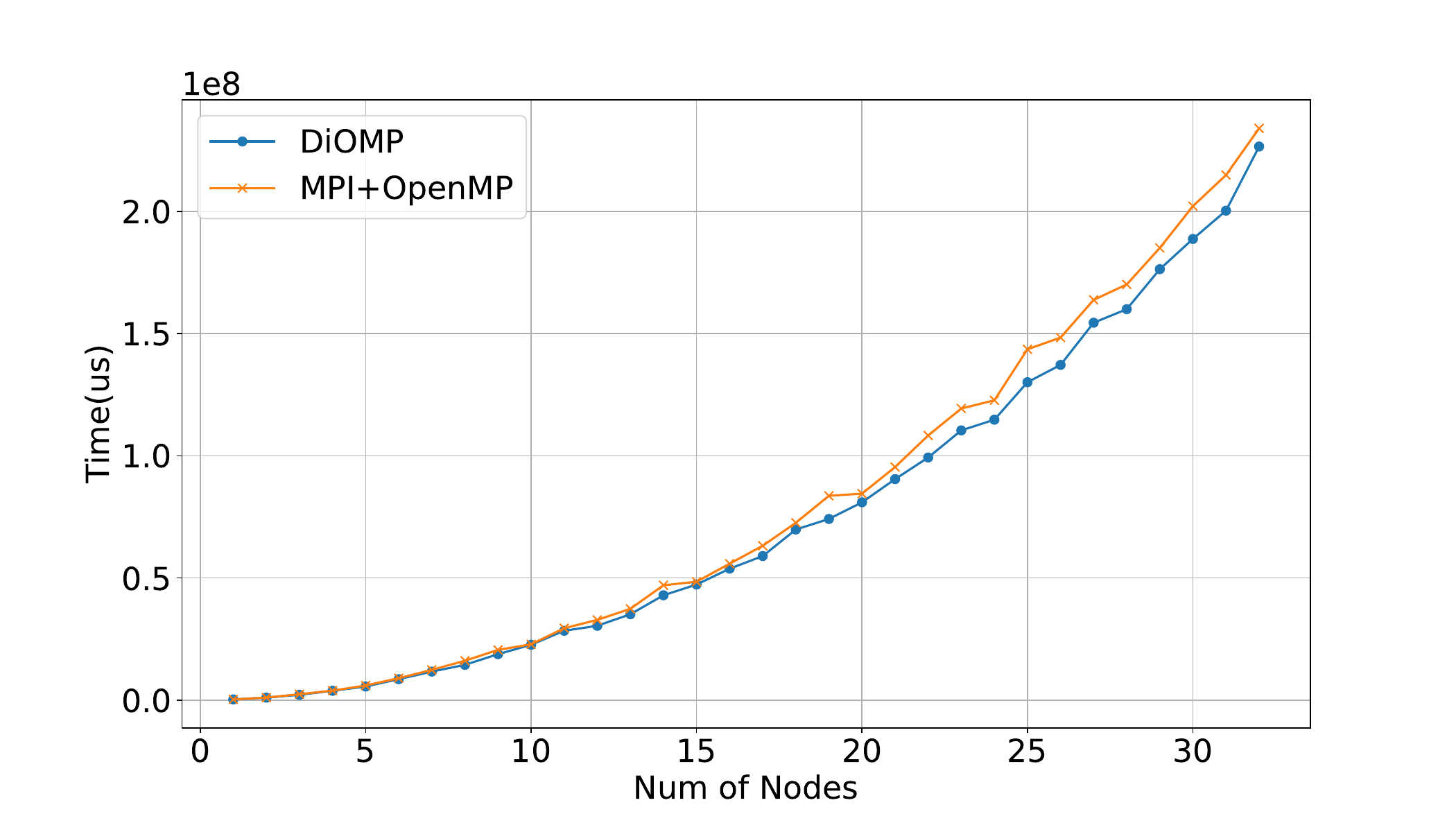}
    \caption{Matrix Multiplication on Ookami}
    \label{fig:mm}
\end{figure}

We subsequently evaluate the ring exchange communication pattern using a mini-application that implements Cannon's algorithm to perform square matrix multiplication, resulting in the product $C
 = A \times B$. Both the MPI version and the DiOMP version of the mini-app incorporate an additional bLoCk stripe for matrix B, enabling the overlap of computation and communication. In this mini-app, as the number of ranks increases, the size of the matrix and the volume of data transferred also increase. 
In this test, the matrix size is $500\times500\times\text{ranks number}$, resulting in a linear increase in computational load. Due to the ring communication pattern employed, the volume of communication increases in squares. \autoref{fig:mm} presents the results of matrix multiplication on the Ookami system using both DiOMP and MPI+OpenMP.

\subsection{Strong Scaling - Minimod}

\begin{table}[!h]
\centering
\caption{Lines of code of MPI+OpenMP verus PGAS+OpenMP}
\label{tab:LoC}
\resizebox{.5\linewidth}{!}{%
\begin{tabular}{cc}
\hline
\multicolumn{1}{l}{\textbf{\begin{tabular}[c]{@{}c@{}}Programming Model\end{tabular}}} & \multicolumn{1}{l}{\textbf{\begin{tabular}[c]{@{}c@{}}Lines of Code\end{tabular}}} \\ \hline
MPI+OpenMP                              & 26                                         \\
PGAS+OpenMP                               & 14                                         \\ \hline
\end{tabular}}
\end{table}

Minimod~\cite{meng2020minimod} is a proxy application designed to simulate the propagation of waves through subsurface models by solving the wave equation in its finite difference discretized form. In this study, we utilize one of the kernels included in Minimod, specifically the acoustic isotropic propagator in a constant-density domain~\cite{qawasmeh2017performance}.

\begin{figure}[!h]
    \centering
    \includegraphics[width=\textwidth]{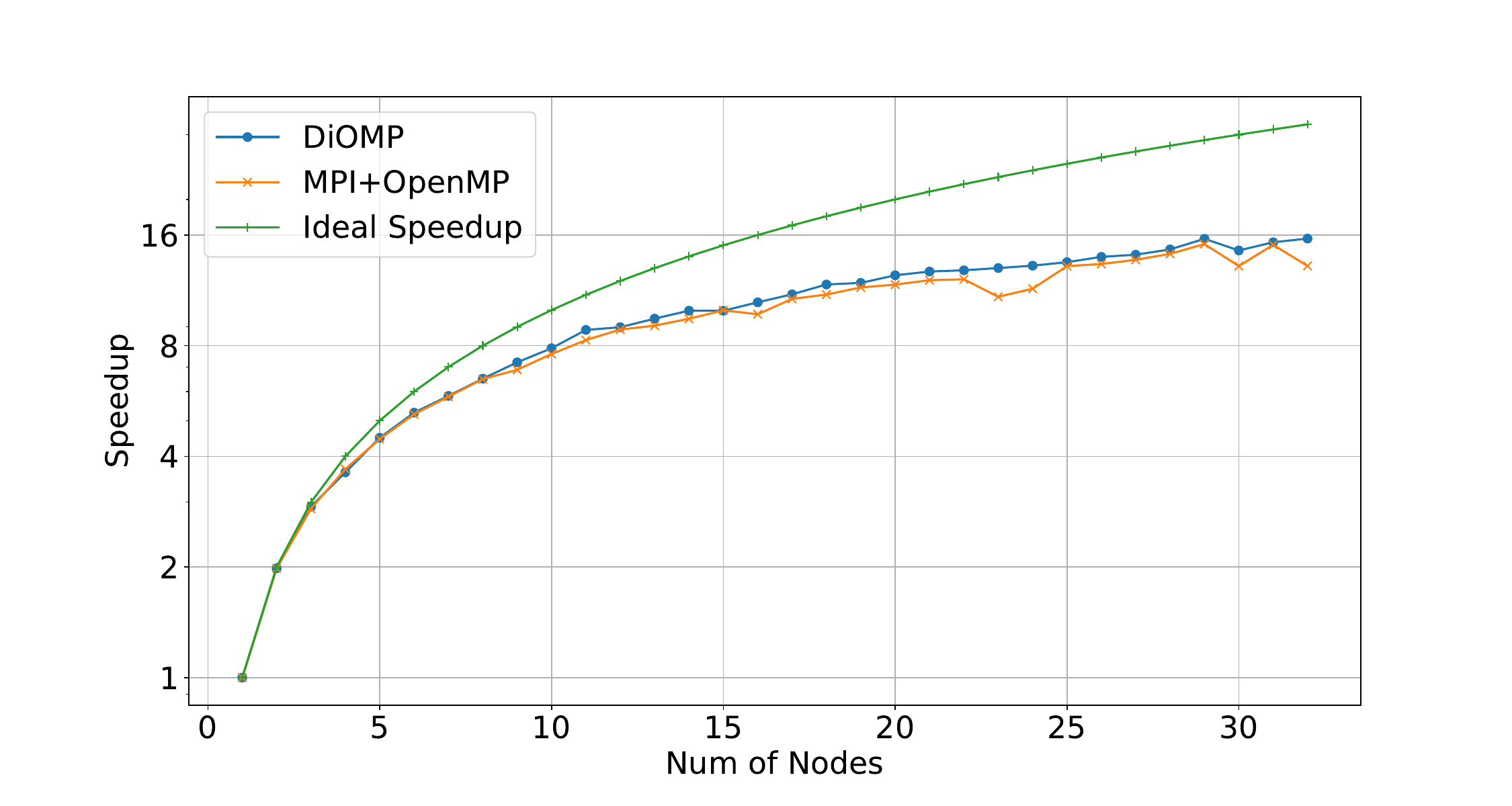}
    \caption{Minimod on Ookami}
    \label{fig:minimod}
\end{figure}

Minimod supports multi-device OpenMP offloading using \texttt{target} regions encapsulated within OpenMP tasks and exhibits strong-scaling characteristics~\cite{raut2020evaluating}. We ported the multi-GPU version of Minimod to versions using MPI+OpenMP and DiOMP. In these versions, the original GPUs device numbers are treated as ranks, with data exchanges being handled through PGAS or MPI. Remarkably, the MPI+ OpenMP version LoCs required for communication are significantly larger than those for the DiOMP version as shown in ~\autoref{lst:mpi} and~\autoref{lst:omp}. 
In~\autoref{lst:mpi}, since MPI uses two-sided communication, both the sender and receiver need to be involved in the data transmission process, in order to minimize the waiting time, we set up \texttt{MPI\_Request} arrays for both sides of the transmission to ensure the synchronization of information.
In~\autoref{lst:omp}, since DiOMP uses windowless one-sided communication, the data sender only needs to put the data to the target rank. The \texttt{ompx\_waitALLRMA} will wait for all data to be received completely before executing the code below.
The specific values can be referenced in~\autoref{tab:LoC}.
For tests in \autoref{fig:minimod}, the grid size is \(1000^3\) and \(1000\) time steps. We conducted evaluations on the Ookami system using 1 to 32 nodes. \autoref{fig:minimod} shows the results of Minimod running on Ookami using both DiOMP and MPI+OpenMP versions. We observed excellent strong scalability.
It is clear that in the majority of cases, DiOMP demonstrated either comparable or superior performance to MPI+OpenMP.

\begin{figure}[H]
\begin{lstlisting}[caption={Minimod - MPI}, label=lst:mpi]
MPI_Request requests[4*nranks];
int req_cnts[nranks];
memset(req_cnts, 0, nranks*sizeof(int));
for (int r=0; r<nranks; r++) {
    RANK_XMIN_XMAX(r,gxmin,gxmax);
    if (rank == r) {
        if (r != 0) {
            rc = MPI_Isend(..., &requests[req_cnts[r]++]);
        }
        if (r != nranks-1) {
            rc =  MPI_Isend(..., &requests[req_cnts[r]++]);
        }
    }
    if (rank == r-1) {
        rc = MPI_Irecv(..., &requests[req_cnts[r]++]);
    }
    if (rank == r+1) {
        rc = MPI_Irecv(..., &requests[req_cnts[r]++]);
}}
for (int r=0; r<nranks; r++) {
    if (req_cnts[r] > 0) {
        MPI_Waitall(req_cnts[r], requests, MPI_STATUSES_IGNORE);
}}        
\end{lstlisting}
\end{figure}

\begin{figure}[H]
\begin{lstlisting}[caption={Minimod - DiOMP}, label=lst:omp]
for (int r = 0; r < nranks; ++r) {
    llint gxmin, gxmax;
    RANK_XMIN_XMAX(r,gxmin,gxmax);
    if (r != 0) {
        if(rank == r){
            ompx_put(...);
    }}
    if (r != nranks-1) {
        if(rank == r){
            ompx_put(...);
}}}
ompx_waitALLRMA();     
\end{lstlisting}
\end{figure}
\section{Conclusion and Future Work}

In conclusion, this paper introduces \textbf{DiOMP}, an extension of OpenMP utilizing the PGAS distributed model. DiOMP leverages LLVM/OpenMP and GASNet-EX to offer a portable, scalable, and high-performance solution for parallel programming across diverse architectures.
We hope that DiOMP can become an important extension of OpenMP and eventually become part of the OpenMP specification. Based on the current experimental results, DiOMP achieves competitive performance against the legacy MPI+X approach. The PGAS-based Distributed OpenMP model has the potential to replace the traditional MPI+OpenMP hybrid programming approach in many scenarios.


Looking ahead, we aim to further expand the usability of DiOMP, particularly with respect to OpenMP target offloading, including support for accelerators like GPUs, and managing OpenMP task dependencies through active message. We also intend to apply the PGAS-based Distributed OpenMP model to real-world scientific applications and study its productivity and performance in comparison with other PGAS approaches and the MPI+OpenMP hybrid model.

\section*{Acknowledgements}
We would like to thank TotalEnergies E\&P Research and Technologies US for their support of this work. Our gratitude also extends to Alice Koniges from the University of Hawaii for providing access to the NERSC Perlmutter system. 

Additionally, we acknowledge to thank Stony Brook Research Computing and Cyberinfrastructure, and the Institute for Advanced Computational Science at Stony Brook University for access to the innovative high-performance Ookami computing system, which was made possible by a \$5M National Science Foundation grant (\#1927880). This research also used resources of the National Energy Research Scientific Computing Center, which is supported by the Office of Science of the U.S. Department of Energy under Contract No. DE-AC02-05CH11231. 

%

%

%
%
%
%

\bibliographystyle{splncs04}
\bibliography{iwomp}

\end{document}